\documentclass[twocolumn]{autart}

\usepackage[longnamesfirst]{natbib}
\usepackage{graphicx}
\usepackage{amsmath}
\usepackage{amssymb}
\usepackage{xcolor}
\usepackage{cases}
\usepackage[noadjust]{cite}
\usepackage[shortlabels]{enumitem}
\usepackage{comment}

\newtheorem{lem1}{\bf Lemma}
\newtheorem{assmpt1}{\bf Assumption}
\newtheorem{rem1}{\bf Remark}
\newtheorem{cor1}{\bf Corollary}

\newenvironment{assmpt}{\begin{assmpt1}}{\end{assmpt1}}

\newcommand{\R}{\ensuremath{{\mathbb R}}}
\newcommand{\Z}{\ensuremath{{\mathbb Z}}}
\newcommand{\N}{\ensuremath{{\mathbb N}}}
\newcommand{\NN}{{\mathcal N}}
\newcommand{\GG}{{\mathcal G}}
\newcommand{\EE}{{\mathcal E}}

\newcommand{\XX}{{\mathcal X}}

\usepackage{chngcntr}
\usepackage{apptools} 
\AtAppendix{\counterwithin{lem1}{section}}

\DeclareMathAlphabet{\mathmybb}{U}{bbold}{m}{n}
\newcommand{\1}{\mathmybb{1}}
\newcommand{\0}{\mathmybb{0}}

\begin{document}
	
	\begin{frontmatter}
		
		\title{A Design Method of Distributed Algorithms via Discrete-time Blended Dynamics Theorem\thanksref{footnoteinfo}} 
		\thanks[footnoteinfo]{This work was supported by the National Research Foundation of Korea(NRF) grant funded by the Korea government(Ministry of Science and ICT) (No. RS-2022-00165417).}
		
		\author[Korea]{Jeong Woo Kim}\ead{jwkim@cdsl.kr}, 
		\author[UK]{Jin Gyu Lee}\ead{jingyulee.scholar@gmail.com}, 
		\author[Korea]{Donggil Lee}\ead{dglee@cdsl.kr}, and
		\author[Korea]{Hyungbo Shim}\ead{hshim@snu.ac.kr}
		
		\address[Korea]{ASRI, Department of Electrical and Computer Engineering, Seoul National University, Seoul, Korea} \address[UK]{CAP Research Group, Department of Electrical and Electronic Engineering, Imperial College London, United Kingdom}
				
				
		\begin{abstract}
			We develop a discrete-time version of the blended dynamics theorem for the use of designing distributed computation algorithms. 
			The blended dynamics theorem enables to predict the behavior of heterogeneous multi-agent systems. 
			Therefore, once we get a blended dynamics for a particular computational task, design idea of node dynamics for individual heterogeneous agents can easily occur.
			In the continuous-time case, prediction by blended dynamics was enabled by high coupling gain among neighboring agents. 
			In the discrete-time case, we propose an equivalent action, which we call multi-step coupling in this paper.
			Compared to the continuous-time case, the blended dynamics can have more variety depending on the coupling matrix.
			This benefit is demonstrated with three applications; distributed estimation of network size, distributed computation of the PageRank, and distributed computation of the degree sequence of a graph, which correspond to the coupling by doubly-stochastic, column-stochastic, and row-stochastic matrices, respectively. 
		\end{abstract}
		
	\end{frontmatter}

	\section{Introduction} \label{sect:intro}
		
	Over the past decades, many distributed algorithms have been actively studied for their benefits.
	The benefits include lessened computational burden of one node as the burden is distributed over many nodes in the network, 
	improved reliability against faults as a fault on one node can be compensated by redundancy of many nodes, and preserved privacy as private information need not be transferred to a central node for computation.
	Examples that enjoy the aforementioned benefits include distributed optimization \citep{AN09,nedic2018} and distributed computation of PageRank \citep{HI10,HI12}.	
	
	On the other hand, constructive design methods for general distributed algorithms, except the distributed optimization, are not well developed yet.
	One potential approach towards the constructive design is the {\em blended dynamics} approach \citep{JL20}, which is inspired by \citep{JK16,panteley2017}.
	This approach is based on the blended dynamics theorem, which, briefly speaking, asserts the following.
	Consider a multi-agent system
	\begin{align}\label{eq:contMAS}
		\dot{\mathsf{x}}_i = \mathsf{f}_i(\mathsf{t},\mathsf{x}_i) + \kappa \sum_{j \in \NN_i} (\mathsf{x}_j-\mathsf{x}_i), \quad i \in \NN,
	\end{align} 
	where $\NN := \{1,\cdots,N\}$ is the set of node indices, and $\NN_i$ is the index set of nodes that send information to node~$i$. 
	The individual node dynamics is represented by $\dot{\mathsf{x}}_i = \mathsf{f}_i(\mathsf{t},\mathsf{x}_i)$, which is coupled with neighboring nodes by the coupling term $\sum_{j \in \NN_i} (\mathsf{x}_j-\mathsf{x}_i)$ with a common coupling gain $\kappa$.
	Then, under the assumption that the communication graph is undirected and connected, every agent in \eqref{eq:contMAS} behaves like the blended dynamics defined as
	\begin{align}\label{eq:contBD}
		\dot{\mathsf{s}}(\mathsf{t}) = \frac{1}{N}\sum_{i=1}^N \mathsf{f}_i(\mathsf{t},\mathsf{s}(\mathsf{t})) 
	\end{align}
	if the coupling gain $\kappa$ is sufficiently large and if the blended dynamics is contractive, i.e., incrementally stable \citep{JK16}.
	More precisely, for any $\epsilon > 0$, there exists $\kappa^{\min}$ such that, if $\kappa > \kappa^{\min}$,
	\begin{equation}\label{eq:recall}
		\limsup_{\mathsf{t} \to \infty} |\mathsf{x}_i(\mathsf{t}) - \mathsf{s}(\mathsf{t})| \le \epsilon, \qquad \forall i \in \NN.
	\end{equation}
	Since the blended dynamics is a simple average of individual node dynamics, it has been utilized as a design tool for many distributed algorithms; that is, one designs a desired algorithm as the blended dynamics \eqref{eq:contBD} first, and then, splits it into different node dynamics \eqref{eq:contMAS} with couplings.
	This philosophy has been successfully employed in many applications such as distributed economic power dispatch problem \citep{HY18}, distributed state estimator \citep{TK19}, secure estimation by distributed median computation \citep{lee2020fully}, distributed least square solver \citep{lee2019distributed}, distributed optimization without convexity of each node \citep{SL22}, and decentralized controller design \citep{kim2020decentralized}.
	See \citep{JG22} for more comprehensive summary of these applications.
	The distributed algorithms designed by the blended dynamics theorem does not require each node dynamics to be stable, as long as their average (i.e., the blended dynamics) is contractive, which yields flexibility of the design.
	Moreover, as long as the blended dynamics remains contractive, a new node can join the network or an existing node can leave the network during the operation, which is called as a plug-and-play feature. 
	This is because the designed distributed algorithms are initialization-free \citep{JL20}.
	
	While all the above results are in the continuous-time domain,	it is however required to implement the designed algorithm in the discrete-time domain so that it operates on digital devices in practice.
	A naive idea is to use simple discretization methods such as forward difference. 
	For example, a discretized model of \eqref{eq:contMAS} becomes
	\begin{multline}\label{eq:discretizedContMAS}
		\mathsf{x}_i(\mathsf{t}+\Delta_\mathsf{t}) = \mathsf{x}_i(\mathsf{t})+\Delta_\mathsf{t} \mathsf{f}_i(\mathsf{t},\mathsf{x}_i(\mathsf{t})) \\
		+ \kappa \Delta_\mathsf{t} \sum_{j\in\NN_i} (\mathsf{x}_j(\mathsf{t})-\mathsf{x}_i(\mathsf{t})),
	\end{multline}
	where $\Delta_\mathsf{t}$ is the sampling time. 
	In the continuous-time case, we recall that arbitrarily small error between $\mathsf{x}_i$ and $\mathsf{s}$ in \eqref{eq:recall} can be obtained by increasing the coupling gain $\kappa$, so that an emergent behavior of the multi-agent system arises with strong coupling.
	However, in the discrete-time case of \eqref{eq:discretizedContMAS}, increasing $\kappa$ unboundedly yields instability of the network unless $\Delta_\mathsf{t}$ is decreased with the same ratio\footnote{One can verify it with, e.g., $\mathsf{f}_i(\mathsf{t},\mathsf{x}_i) = -\mathsf{x}_i$ so that $\dot{\mathsf{x}} = \left\{ (1 - \Delta_\mathsf{t}) I - \kappa \Delta_\mathsf{t} \mathcal{L} \right\} \mathsf{x}$ where $\mathsf{x} = [\mathsf{x}_1,\dots,\mathsf{x}_N]^T$ and $\mathcal{L}$ is the Laplacian matrix of a connected graph. With $\kappa$ sufficiently large, some eigenvalues of the system matrix lie outside of the unit circle unless $\kappa\Delta_\mathsf{t}$ remains small.}.
	Therefore, the discrete-time algorithm \eqref{eq:discretizedContMAS} cannot be a discrete-time version of the blended dynamics approach.
	
	In this paper, we propose a new form of a multi-agent system (which is given by \eqref{eq:maineq} in Section \ref{sect:main}).
	We note that the meaning of using a large coupling gain $\kappa$ in the continuous-time case of \eqref{eq:contMAS} is that consensus is taken more care of than the progress through the node dynamics.
	Based on the observation, and motivated by \citep{wang2019distributed}, the proposed form repeats a weighted averaging action many times before progressing through the node dynamics, which we call `multi-step coupling.'
	
	This approach maintains the advantages of the continuous-time case, such as the plug-and-play operation, and that the individual node dynamics need not be stable as long as the blended dynamics is stable, which we do not repeat in this paper but refer the reader to \citep{JK16,JL20,kim2020decentralized}.
	
	Moreover, while the continuous-time approach predicted collective synchronization behavior of the multi-agent system in \citep{JK16,JL20}, this discrete-time approach estimates not only emergent but also individually scaled behavior, i.e., each agent behaves similarly to the solution of the blended dynamics with an agent-wise scaling factor. 
	For example, in Section~\ref{subSect:PRCp}, we will introduce an application example where each node estimates its relative importance which is possibly agent-wise different so that overall nodes are not synchronized in the network.

	\subsection{Notation and useful facts}
	
	We use the following convention in this paper.
	$\0_N$ and $\1_N$ denote the column vectors of size $N$ consisting of all zeros and ones, respectively. 
	A positive vector implies a vector whose elements are all positive.
	The 2-norm of a vector and the induced 2-norm of a matrix are denoted by $\Vert \cdot \Vert$. 
	For any square matrix $\mathsf{M}$, $\rho(\mathsf{M})$ denotes the spectral radius of $\mathsf{M}$.
	The operation defined by the symbol $\otimes$ is Kronecker product, which has the properties that, given the matrices $\mathsf{A,B,C}$ and $\mathsf{D}$ of appropriate dimensions, $(\mathsf{A} \otimes \mathsf{B})(\mathsf{C}\otimes \mathsf{D}) = (\mathsf{AC})\otimes(\mathsf{BD})$ and $\| \mathsf{A}\otimes\mathsf{B} \| = \|\mathsf{A}\| \|\mathsf{B}\|$.
	See, e.g., \citep{bernstein2009matrix} for more about the Kronecker product.
	For notational convenience, we use the convention $\mathsf{M}_{\otimes n} := \mathsf{M}\otimes I_n$ for any size of matrix $\mathsf{M}$. 
	
	A communication network of $N$ agents can be represented by a directed graph $\GG=(\NN,\EE)$ where $\NN=\left\{1,2,\dots,N\right\}$ is the node set, and $\EE\subset\NN\times\NN$ is the edge set of ordered pairs of nodes. If agent $j$ sends information to agent $i$,
	then we say that the node $j$ is connected to the node $i$ by an edge $(j,i)\in\EE$. 
	A graph is said to be undirected if $(j,i)\in\EE$ implies $(i,j)\in\EE$. 
	For a directed graph, in-neighbors of node $i$ are represented by $\NN_i=\{j\in\NN|(j,i)\in\EE\}$ and in-degree of node $i$ is denoted by $d_i:=|\NN_i|$. 
	On the contrary, out-neighbors of node $i$ are represented by $\NN_i^\text{out}:=\{j\in\NN|(i,j)\in\EE\}$ and out-degree of node $i$ is denoted by $d_i^\text{out}:=|\NN_i^\text{out}|$. 
	A path of length $L$ from node $i$ to node $j$ is a sequence $(i_0,i_1,\ldots,i_L)$ such that $i_0=i$, $i_L=j$, $(i_l,i_{l+1})\in\EE$ for any $l\in\{0,\ldots,L-1\}$, and every $i_l$ is distinct. 
	A directed graph is said to be strongly connected if there exists a path from any node to any other node. 
	Similarly, an undirected graph is said to be connected if there exists a path between any two nodes. 
	A cycle is a path that starts and ends at the same node and any node does not appear more than once in it.	
	A strongly connected directed graph is said to be periodic if there exists period $\mathsf{p}>1$ that divides the length of every cycle in the graph, otherwise the graph is said to be aperiodic. 
	Any directed graph having at least one self-connection is aperiodic. 
	For a directed graph $\GG$, its { adjacency matrix} $\mathcal{A}=[\alpha_{ij}]\in\R^{N\times N}$ is defined as $\alpha_{ij}>0$ if $(j,i)\in\EE$ and $\alpha_{ij}=0$ otherwise. 
	If every positive component of $\mathcal{A}$ is 1, then $\mathcal{A}$ is called a binary adjacency matrix.
	Meanwhile, given a non-negative square matrix $\mathsf{M}\in\R^{N\times N}$, its associated directed graph is defined as the directed graph whose adjacency matrix is $\mathsf{M}$. 
	A non-negative square matrix $\mathsf{M}\in\R^{N \times N}$ is said to be {\em primitive} if there exists $\mathsf{n}\in\N$ such that $\mathsf{M}^\mathsf{n}$ is {\em positive}, i.e., every component of $\mathsf{M}^\mathsf{n}$ is positive. 
	The primitive property of any non-negative square matrix can be verified by its associated graph as the following lemma.
	
	\begin{lem1}[{\citet[Theorem 4.7]{FB20}}]\label{lemPrimitiveGraph}
		For a directed graph $\mathcal{G}$, its adjacency matrix $\mathcal{A}$, and its binary adjacency matrix $A$, the following statements are equivalent: (a) $\mathcal{G}$ is strongly connected and aperiodic, (b) $\mathcal{A}$ is primitive, and (c) $A$ is primitive.
	\end{lem1}	
	
	\begin{lem1}[Perron-Frobenius theorem]\label{lemPFthm}
		If a non-negative matrix $\mathsf{M}\in\R^{N\times N}$ is primitive, then the following statements hold:
		\begin{enumerate}[1)]
			\item There exists a simple eigenvalue $\lambda>0$ (called Perron-Frobenius eigenvalue) of $\mathsf{M}$ such that $\lambda >$ $|\sigma|$ for any other eigenvalue $\sigma$ of $\mathsf{M}$.
			\item The right and left eigenvectors of $\lambda$ are positive.
		\end{enumerate}
	\end{lem1}

	\section{Discrete-time Blended Dynamics Theorem} \label{sect:main}	
	
	For a discrete-time version of the blended dynamics theorem, we propose the following discrete-time algorithm: for each agent $i \in \NN$, 
	\begin{subnumcases}{x_i[t_{k+1}]= \label{eq:maineq}}
		f_i(t_k, x_i[t_k]), & \small $\text{if } k=0,$ \label{eq:heteroDynamics} \\
		\sum_{j \in \NN_i \cup \{i\}} w_{ij} x_j[t_k],& \small $\text{if } k=1,\ldots,K-1,$ \label{eq:couplingDynamics}
	\end{subnumcases}
	where $x_i \in \R^n$ is the state, the function $f_i: \mathbb{Z} \times \R^n \rightarrow \R^n$ is continuously differentiable and represents the time-varying heterogeneous node dynamics \eqref{eq:heteroDynamics}, and the coefficient $w_{ij}$, called {\em coupling weight}, determines the behavior of the coupling dynamics \eqref{eq:couplingDynamics}.
	Here, $t_k$ is the symbol defined by
	\begin{align}\label{eq:fracTimeIndex}
		t_k = t + \frac{k}{K}, 
	\end{align}
	where $K \in \N$, and we call $t_k$ by {\em fractional discrete-time index}.
	In particular, we call $t \in \Z$ as {\em integer count} and $k \in \N$ as {\em fraction count}.
	The fraction count $k$ varies from 0 to $K-1$.
	Keeping in mind that $t_K=t+K/K=(t+1)+0/K=(t+1)_0$, we see that the fractional discrete-time $t_k$ advances as $0_0, 0_1, \cdots, 0_{K-1}, 1_0, 1_1, \cdots$.
	The time $t_0$ will often be written as $t$ for convenience.
	The fractional discrete-time has nothing to do with real time, and can be implemented in practice just as a sequential order in an algorithm.
	
	We will choose $K$ sufficiently large, which determines how many times the coupling dynamics \eqref{eq:couplingDynamics} is executed before the next node dynamics \eqref{eq:heteroDynamics} is executed.
	It will be shown that, in this way, the effect of {\em strong coupling} $\kappa$ in the continuous-time blended dynamics theorem can be similarly reflected in discrete-time.
	To emphasize the difference, we call this type of coupling in \eqref{eq:maineq} by {\em multi-step coupling}.
	
	The coupling weights $w_{ij}$ in \eqref{eq:couplingDynamics} have the property:
	\begin{align}\label{eq:defOfWeight}
		w_{ij}
		\begin{cases}
			>0, &j\in\NN_i\cup\{i\},\\
			=0, &\text{otherwise}.
		\end{cases}
	\end{align}
	Now, we assume that the matrix $W := [w_{ij}] \in \R^{N\times N}$, which we call a {\em weight matrix}, satisfies the following.
	
	\begin{assmpt} \label{assW}
		The spectral radius of $W$ is 1. 
	\end{assmpt}
	
	Note that the communication protocols in many discrete-time multi-agent systems in the literature have the form of linear combination like in \eqref{eq:couplingDynamics} and their weight matrices satisfy Assumption \ref{assW}. 
	Examples include \citep{HI10, HI12, JL14, RS07, WR05}, in which the weight matrices are given by stochastic matrices whose spectral radius is 1.
	
	Meanwhile, the communication network under consideration is represented by the directed graph $\GG$, which does not have self-connection at any node by definition, and we assume the following.
	
	\begin{assmpt} \label{assGraph}
		The network $\GG$ is strongly connected.
	\end{assmpt}
	
	Then, under Assumptions \ref{assW} and \ref{assGraph}, the following is well-known (but we put its proof for readers' convenience).
		
	\begin{lem1} \label{lemPQ}
		Let $\lambda_i$, $i \in \NN$, be the eigenvalues of $W$ such that $|\lambda_1| \ge |\lambda_2| \ge \ldots \ge |\lambda_N|$. 
		Under Assumptions \ref{assW} and \ref{assGraph}, $\lambda_1 = 1$, $\lambda_1 > |\lambda_j|$ for all $j=2,\ldots,N$, and there exist positive vectors $p,q \in \R^N$ such that
		\begin{align}\label{eq:defOfEigenvectors}
			Wp=p, \quad q^\top W =q^\top, \quad q^\top p=1.
		\end{align}
	\end{lem1}
	
	\begin{pf}
		From \eqref{eq:defOfWeight}, the associated graph of $W$ not only contains all edges of $\GG$ but also has a self-connection for every node because all diagonal entries of $W$ are positive. 
		Thus, the associated graph is aperiodic as well as strongly connected.
		This implies that $W$ is primitive by Lemma~\ref{lemPrimitiveGraph}, and, by Assumption~\ref{assW} and Lemma~\ref{lemPFthm}, $W$ has the simple Perron-Frobenius eigenvalue 1, i.e., $\lambda_1=1$ and $\lambda_1 > |\lambda_j|$ for all $j = 2, \dots, N$, with positive right and left eigenvectors $p$ and $q$, respectively.
		Scaling $p$ and $q$ yields that $q^\top p = 1$.	$\hfill\blacksquare$
	\end{pf}
	
	With $p$ and $q$ from Lemma \ref{lemPQ} at hand, we now introduce {\em discrete-time blended dynamics}, which is defined as a {\em weighted} average of node dynamics:
	\begin{align}\label{eq:blendedDynamics0}
		s[t + 1]&=\sum_{i=1}^{N} q_i f_i(t, p_i s[t]) =: f_s(t,s[t]) \quad \in \R^n
	\end{align}
	where $t$ is the integer count of the fractional time (i.e., $t=t_0$).
	In particular, we assume the blended dynamics is stable in the sense of \citep{LW98,tran2018convergence} as follows.
	
	\begin{assmpt} \label{assStability}
		The blended dynamics \eqref{eq:blendedDynamics0} is {\em contractive}; i.e., there exist a (symmetric) positive definite matrix $H \in \R^{n \times n}$ and a positive constant $\gamma < 1$ such that 
		$$\frac{\partial f_s}{\partial s}(t,s)^\top H^2 \frac{\partial f_s}{\partial s}(t,s) \le \gamma H^2, \quad \forall s \in \mathbb{R}^n, t \in \Z.$$
	\end{assmpt}
	
	\begin{rem1}
		Assumption \ref{assStability} does not ask each node dynamics $x_i[t+1]=f_i(t,x_i[t])$ to be stable.
		Rather it allows unstable node dynamics whose instability can be compensated by other node dynamics so that the blended dynamics becomes stable in the sense of Assumption~\ref{assStability}. 
		For example, when there are four agents with $f_1(t,x) = f_2(t,x) = 0.1 x$ and $f_3(t,x) = f_4(t,x) = 1.5x$, the agents 1 and 2 have stable node dynamics while the agents 3 and 4 have unstable ones. 
		If the weight matrix has the vectors $p = \1_4$ and $q = \1_4/4$, then Assumption~\ref{assStability} holds because $f_s(s) = 0.8s$.
	\end{rem1}	
	
	We will see that the blended dynamics \eqref{eq:blendedDynamics0} allows to predict the behavior of \eqref{eq:maineq} when $K$ is large.
	To make the prediction effective from any initial conditions globally in the state-space, we impose the following assumption.
	
	\begin{assmpt} \label{assFi}
		The function $f_i(t,x)$ is uniformly bounded in $t$ and globally Lipschitz with respect to $x$ uniformly in $t$: i.e., $\exists$ a non-decreasing continuous function $M:\mathbb{R} \rightarrow \mathbb{R}$ and a constant $L\ge0$ such that, $\forall x,y \in \mathbb{R}^n$, $t \in \mathbb{Z}$, and $i \in \mathcal{N}$,
		\begin{gather*}
			\| f_i(t,x) \| \le M\left( \|x\| \right), \\
			\| f_i(t,x)-f_i(t,y) \| \le L \| x - y \|.
		\end{gather*}
	\end{assmpt}
	
	\begin{thm} \label{thm1}
		Under Assumptions \ref{assW}--\ref{assFi}, for any $\epsilon>0$, there exists $K^\mathrm{min}$ such that, for all $K > K^\mathrm{min}$, the solution $x_i$ of \eqref{eq:maineq} and the solution $s$ of \eqref{eq:blendedDynamics0} with arbitrary initial conditions 
		satisfy
		\begin{align}\label{eq:thm1Eq1}
			\limsup_{t\rightarrow\infty} \big\| {x}_i[t]-p_i s[t] \big\| \le \epsilon, \quad \forall i\in\NN.
		\end{align}
		In addition, 
		for each $k \in \{1,2,\ldots,K-1\}$ and $i \in \NN$,
		\begin{equation}\label{eq:comparefraction}
			\limsup_{t \to \infty} \big\| x_i[t_k] - p_i s[t+1] \big\| \le \frac{\epsilon}{2} \left( 1 + \frac{1}{|\lambda_N|^{K-k}} \right) .
		\end{equation}
	\end{thm}
	
	Theorem \ref{thm1} states that, with sufficiently large number of steps for the coupling \eqref{eq:couplingDynamics}, the behavior of node dynamics \eqref{eq:heteroDynamics}, which is represented by the state $x_i$ at the integer count $t$, can be approximately predicted by the solution $s$ of the blended dynamics with the scaling factor $p_i$, and the approximation error can be made arbitrarily small by increasing $K$.
	Moreover, the behavior of $x_i$ over the fraction counts is also bounded with respect to the scaled trajectory of $s$.
		
	\begin{rem1}
		Selection of the eigenvectors $p$ and $q$ as \eqref{eq:defOfEigenvectors} is not unique, but the result of Theorem \ref{thm1} remains the same.
		To see this, we note that different selection of $p'$ and $q'$ from $p$ and $q$, respectively, should satisfy $p'=cp$ and $q'=(1/c)q$ for some $c>0$ because of the Perron-Frobenius theorem (Lemma \ref{lemPFthm}).
		In addition, we note that the new blended dynamics becomes $s'[t+1] = (1/c) \sum_{i=1}^N q_i f_i(t, c p_i s'[t]) =: f_{s'}(t,s'[t])$.
		Comparing it with \eqref{eq:blendedDynamics0}, it is seen that $s'[t] = (1/c)s[t]$, and thus, we have $\| x_i[t] - p'_i s'[t] \| = \| x_i[t] - c p_i (1/c) s[t] \| = \| x_i[t] - p_i s[t] \|$. 
		Finally, it is also seen that the new blended dynamics satisfies Assumption \ref{assStability} because $(\partial f_{s'}/\partial s')(t,s') = (\partial f_s/\partial s)(t,s)$ with $s = cs'$.
	\end{rem1}
	
	\begin{rem1}
		In \eqref{eq:comparefraction}, the state $x_i[t_k]$ is compared not with $s[t]$ but with $s[t+1]$.
		One may find this is natural considering the behavior of the overall system.
		At each integer time $t = t_0$, each $x_i$ obeys the heterogeneous node dynamics \eqref{eq:heteroDynamics}, which potentially updates $x_i[t_1]$ in different directions from the updated $p_i s[t+1]$ (even if $x_i[t_0]$ is close to $p_i s[t]$).
		Instead, repeated execution of \eqref{eq:couplingDynamics} drives $x_i[t_k]$ to $p_i s[t+1]$, which is well reflected in \eqref{eq:comparefraction}.
	\end{rem1}
		
	We now present intuitive explanations for Theorem \ref{thm1}, whose rigorous proof continues in the Appendix.
	For simplicity, define $\bar x := [x_1^\top, \ldots, x_N^\top]^\top\in \R^{nN}$.
	Then, \eqref{eq:maineq} is simply written as
	\begin{equation}\label{eq:realoverall}
		\bar x[t_k] = W_{\otimes n}^{k-1} \begin{bmatrix} f_1(t_0, x_1[t_0]) \\ \vdots \\ f_N(t_0, x_N[t_0]) \end{bmatrix} 
		=: W_{\otimes n}^{k-1} F(t_0, \bar x[t_0]),
	\end{equation}
	for $k=1,\dots,K-1$ and $t \in \Z$, where $W_{\otimes n} = W \otimes I_n$, and, since $t_K = (t+1)_0 = t+1$, we have
	\begin{equation}\label{eq:periodicDynamics}
		\bar x[t+1] = W_{\otimes n}^{K-1} F(t, \bar x[t]).
	\end{equation}
	Similar to \eqref{eq:periodicDynamics}, the blended dynamics \eqref{eq:blendedDynamics0} is written as
	\begin{align}\label{eq:blendedDynamics}
		\begin{split}
			s[t+1] &= \sum_{i=1}^{N} q_i f_i(t, p_i s[t]) = q_{\otimes n}^\top F(t, p_{\otimes n} s[t]).
		\end{split}
	\end{align}
	By Lemma~\ref{lemPQ}, there exist $R, Z\in\R^{N\times (N-1)}$ such that 
	\begin{gather}
		W = \begin{bmatrix}p&R\end{bmatrix} \begin{bmatrix}1&0\\0&\Lambda\end{bmatrix} \begin{bmatrix}q^\top\\Z^\top\end{bmatrix}, \label{eq:decompositionOfW0} \\
		Z^\top R=I_{N-1}, \quad \text{and} \quad Z^\top p = R^\top q = \0_{N-1}, 
	\end{gather}
	where $\Lambda\in\R^{(N-1)\times(N-1)}$ is a matrix whose eigenvalues are $\lambda_2, \ldots, \lambda_N$. 
	With them, we consider the coordinate transformation
	\begin{align}\label{eq:cordTrans}
		\xi = \begin{bmatrix}\xi_1\\\tilde{\xi}\end{bmatrix} = \begin{bmatrix}q_{\otimes n}^\top\\ Z_{\otimes n}^\top \end{bmatrix} \bar x
	\end{align}
	whose inverse is
	$$\bar x = p_{\otimes n}\xi_1 + R_{\otimes n} \tilde{\xi}.$$
	The overall dynamics \eqref{eq:periodicDynamics} at each integer time $t$ becomes
	\begin{align}\label{eq:xiDynamics}
		\begin{split}
			\xi_1[t+1&]= q_{\otimes n}^\top W_{\otimes n}^{K-1} F(t, p_{\otimes n}\xi_1[t] + R_{\otimes n}\tilde{\xi}[t]) \\
			&= q_{\otimes n}^\top F(t, p_{\otimes n}\xi_1[t] + R_{\otimes n}\tilde{\xi}[t]), \\
			\tilde{\xi}[t+1&]= Z_{\otimes n}^\top W_{\otimes n}^{K-1} F(t, p_{\otimes n}\xi_1[t] + R_{\otimes n}\tilde{\xi}[t]) \\
			&= (\Lambda^{K-1} Z^\top)_{\otimes n} F(t, p_{\otimes n}\xi_1[t] + R_{\otimes n}\tilde{\xi}[t]).
		\end{split}
	\end{align}
	Define the error variable $e := \xi_1-s$.
	Then the above dynamics is rewritten by
	\begin{align*}
		e[t+1] &= q_{\otimes n}^\top F(t, p_{\otimes n}(e[t]+s[t]) + R_{\otimes n}\tilde{\xi}[t]) \\
		&\quad - q_{\otimes n}^\top F(t, p_{\otimes n} s[t]), \\
		\tilde{\xi}[t+1] &= (\Lambda^{K-1} Z^\top)_{\otimes n} F(t, p_{\otimes n}(e[t]+s[t]) + R_{\otimes n}\tilde{\xi}[t]).
	\end{align*}
	Since the spectral radius $\rho(\Lambda) = |\lambda_2| < 1$, it may be inferred that $\|\tilde{\xi}\|$ gets small if $K$ is sufficiently large.
	On the other hand, if $\tilde \xi$ happens to be identically zero, then $e[t]$ converges to zero as $t$ tends to infinity, which follows from the following lemma.
	
	\begin{lem1}\label{lemNormIneq}
		Under Assumption \ref{assStability}, 
		\begin{align*}
			\| H \{f_s(t,s_2)-f_s(t,s_1)\} \| &\le \sqrt{\gamma} \| H (s_2 - s_1) \|
		\end{align*}
		for all $t \in \Z$ and $s_1, s_2 \in \R^n$.
	\end{lem1}
	
	In fact, if $\tilde \xi \equiv \0_{n(N-1)}$, then, 
	\begin{align*}
		&\big\| H e[t+1] \big\| \\
		&= \| H \{ q_{\otimes n}^\top F(t, p_{\otimes n}(e[t] + s[t])) - q_{\otimes n}^\top F(t, p_{\otimes n} s[t]) \} \| \\
		&= \| H \{f_s(t, e[t]+s[t]) - f_s(t, s[t])\} \| \le \sqrt{\gamma} \|H e[t]\|.
	\end{align*}
	This implies that $\xi_1$ and $s$ tend to get closer as time goes on. 
	When this happens, $\bar x = p_{\otimes n} \xi_1 + R_{\otimes n} \tilde{\xi}$ tends to $p_{\otimes n} s$, which motivates Theorem \ref{thm1}.
	
	However, $\tilde \xi$ does not become identically zero in general, and the above arguments need rigorous analysis, which is done in the Appendix with a Lyapunov function.
		
	In Theorem \ref{thm1}, the approximation of $x_i[t]$ by $p_i s[t]$ is stated in an asymptotic format, i.e., using $\limsup$.
	If one questions when the approximation becomes effective, the following corollary assures that it can be very early if $K$ is sufficiently large.
	In particular, if the initial conditions of $x_i$ are in a known compact set, then $K^\mathrm{min}$ can be computed (see the proof in the Appendix).
	
	\begin{cor1} \label{thm2}
		Under Assumptions \ref{assW}--\ref{assFi}, for any $\epsilon>0$ and compact set $C \subset \mathbb{R}^{n}$, there exists $K^\mathrm{min}>0$ such that, for all $K > K^\mathrm{min}$, the solution $x_i$ of \eqref{eq:maineq} with $x_i[0] \in C$, and the solution $s$ of \eqref{eq:blendedDynamics0} with $s[1] = \sum_{i=1}^N q_i f_i (0, x_i[0])$
		satisfy
		\begin{align}\label{eq:corineq1}
			\big\| {x}_i[t]-p_i s[t] \big\| \le \epsilon, \quad \forall t \ge 1, \; i \in \NN.
		\end{align}
		In addition, for all $k \in \{1,2,\ldots,K-1\}$, $i \in \NN$, and $t \ge 1$,
		\begin{align}\label{eq:corineq2}
			\big\| x_i[t_k] - p_i s[t+1] \big\| \le \frac{\epsilon}{2} \left( 1 + \frac{1}{|\lambda_N|^{K - k}} \right).
		\end{align}
	\end{cor1}
	
	\begin{rem1}
		In Corollary \ref{thm2}, the solution $s[t]$ of the blended dynamics is initiated not at $t=0$ but at $t=1$.
		One may find this is natural because the state $s'[1] = f_s(0,s'[0])$ with $s'[0] = \sum_{i=1}^N q_i x_i[0]$ (i.e., $s'[1] = \sum_{i=1}^N q_i f_i(0, p_i \sum_{i=1}^N q_i x_i[0])$) can be very different from the considered state $s[1] = \sum_{i=1}^N q_i f_i(0,x_i[0])$.
		In fact, the states $x_j[0_1]$, $j \in \NN$, may be very different from each other, but they converge with sufficiently large $K$ towards $\sum_{i=1}^N q_i x_i[0_1]$ with the scaling factor $p_j$, which $s[1]$ (not $s'[1]$).
	\end{rem1}

	\section{Network Synthesis with Examples} \label{sect:app}	
	
	As mentioned before, the proposed approach is useful as a design method for distributed algorithms by first designing a suitable blended dynamics such that it behaves as desired and then synthesizing each heterogeneity of the multi-agent system such that it has the pre-designed blended dynamics.
	Moreover, comparing with the continuous-time approach, the discrete-time version handles more general protocols as long as the spectral radius of its weight matrix is 1. 
	In fact, many studies on discrete-time multi-agent system including PageRank \citep{HI10, HI12, JL14} or consensus \citep{RS07, WR05} have used the communication protocol which can be represented as a linear combination of agents' state information like \eqref{eq:couplingDynamics} with the weight matrix of unit spectral radius.
	Thus, in this section, some of the protocols are chosen to be considered as the coupling dynamics \eqref{eq:couplingDynamics} and the behaviors of corresponding multi-step coupling systems are illustrated using the results in previous section. 
	Based on this, the design process for each application example is also provided.

	\subsection{Distributed Network Size Estimation with Metropolis-Hastings Coupling}\label{subSect:MHCp}
	
	In this subsection, we assume that the network is undirected and connected, and consider the following Metropolis-Hastings coupling weight $w_{ij}^\text{MH}$ in \citep{VS14}:
	\begin{align*}
		w_{ij}^{\text{MH}} :=
		\begin{cases}
			\displaystyle \frac{1-\mu}{\max\left\{d_i,d_j\right\}}, &  (j,i)\in\mathcal{E} \text{ and } i\neq j,\\
			0, & (j,i)\notin\mathcal{E} \text{ and } i\neq j,\\
			1-\displaystyle \sum_{l\neq i}w_{il}^\text{MH}, & i=j,
		\end{cases}
	\end{align*}
	where $\mu\in(0,1)$. 
	Note that $w_{ij}^\text{MH}$ depends on both $d_i$ and $d_j$, so that each agent should additionally exchange its degree information with neighbors to update its coupling weights online, and this will enable the plug-and-play operation to the multi-step coupling framework.
	
	It can be easily seen that $w_{ij}^\text{MH}$ satisfies \eqref{eq:defOfWeight} and Assumption \ref{assW} holds for $W^{\text{MH}}:=[w_{ij}^{\text{MH}}]$ because it is doubly-stochastic. 
	Hence, we can choose $p=\1_N$ and $q=(1/N)\1_N$ from Lemma~\ref{lemPQ}.
	Then, the blended dynamics is obtained as a simple average of $f_i$s as follows:
	\begin{align}\label{eq:mhBlendedDynamics}
		\begin{split}
			s[t+1] = \frac{1}{N} \sum_{i=1}^{N} f_i(t,s[t]).
		\end{split}
	\end{align}
	Since all $p_i$s are chosen evenly as 1, by Theorem \ref{thm1} or Corollary \ref{thm2}, the behavior of every trajectory ${x}_i[t]$ is approximately synchronized to the solution $s$ of the blended dynamics \eqref{eq:mhBlendedDynamics}.
	It should be emphasized that this collective synchronized behavior comes from $p=\1_N$ and this choice of $p$ is always possible for any row-stochastic (not necessarily to be doubly-stochastic) weight matrices. 
	In Section~\ref{subSect:AvgCp}, we will consider the row-stochastic weight matrix whose column-sums are not 1. 
	
	As an application example for the Metropolis-Hastings coupling, we can design a distributed algorithm for network size estimation as follows. 
	In fact, many distributed algorithms such as \citep{HI12,AN09} are often assumed to know the network size $N$. 
	The insight of the proposed algorithm is to make its blended dynamics converge to $N$.
	For example, if the blended dynamics is designed as the following scalar dynamics
	\begin{align} \label{eq:appMH_BlendedDynamics}
		\begin{split}
			s[t+1]=\left(1-\frac{1}{N}\right)s[t]+1,
		\end{split}
	\end{align}
	such that it has the stable equilibrium point at $N$, then each state $x_i[t]$ will also approach to $N$ under the Metropolis-Hastings coupling. 
	Thus, by increasing $K$ until the synchronization error $\epsilon$ in \eqref{eq:thm1Eq1} or \eqref{eq:corineq1} gets smaller than 0.5, each agent can find the exact network size through the round-off to the nearest integer. 
	
	One idea to design the heterogeneous dynamics $f_i$s whose average becomes \eqref{eq:appMH_BlendedDynamics} comes from
	\begin{align*}
		\frac{1}{N} \left\{ \bigg( 1 \bigg) + \bigg( \sum_{i=2}^N (s[t]+1) \bigg) \right\}=\left(1-\frac{1}{N}\right)s[t]+1,
	\end{align*}
	this is, one agent has $f_i = 1$ and all others have $f_i(s) = s+1$.
	For this, we intentionally add one specific node which does not leave the network during the operation of the algorithm. 
	Without loss of generality, let an index of this node be 1 and it runs the following dynamics:
	\begin{align*}
		x_1[t_{k+1}]=
		\begin{cases}
			1, & \text{if }k=0,\\
			\sum_{j\in\NN_1\cup\{1\}} w_{1j}^\text{MH} x_j[t_{k}], & \text{otherwise}.
		\end{cases}
	\end{align*}
	On the other hand, all the other nodes of $i=2,\ldots,N$ run the following dynamics:
	\begin{align*}
		x_i[t_{k+1}]=
		\begin{cases}
			x_i[t_{k}]+1, & \text{if }k=0,\\
			\sum_{j\in\NN_i\cup\{i\}} w_{ij}^\text{MH} x_j[t_{k}], & \text{otherwise}.
		\end{cases}
	\end{align*}
	Note that, even though the individual dynamics of $(N-1)$ nodes for $i=2,\ldots,N$ are (marginally) unstable, the overall networked system becomes stable and the trajectories of individual agent approach close to $N$ (less than the distance of 0.5 with sufficiently large $K$).
	In addition, this distributed algorithm can be applied even when some agents might join or leave the network during the process of the algorithm, because it does not rely on the initial condition of agents. This idea is motivated by \citep{DL19} which proposed continuous-time distributed network size estimation algorithm.

	\subsection{Initialization-free Distributed PageRank Computation with PageRank Coupling}\label{subSect:PRCp}
	
	Now, we turn our attention to a different type of the weight matrix whose column-sums are all one. 
	In this subsection, we consider the multi-step coupling framework whose coupling dynamics is the following iterative power method of PageRank \citep{SB98}:
	\begin{align}\label{eq:prDynamics}
		x_i[t_{k+1}]=m x_i[t_{k}] + (1-m) \displaystyle\sum_{j\in\NN_i} \frac{x_j[t_{k}]}{d_j^\text{out}},
	\end{align}
	where $x_i\in\R$ is the state, the parameter $m\in(0,1)$ is typically chosen as 0.15, $\NN_i$ is the in-neighbors of node $i$, and $d_j^\text{out}$ is the out-degree of node $j$. 
	In fact, PageRank score provides an information on relative importance of each node in the network, so it has been widely utilized in diverse areas such as informatics \citep{PC07}, bibliometrics \citep{XL05}, and biology \citep{NZ12}. 
		
	Then, the coupling weight $w_{ij}^\text{PR}$ is given by
	\begin{align*}
		w_{ij}^\text{PR}=
		\begin{cases}
			m, & i=j,\\
			(1-m) \displaystyle\frac{a_{ij}}{d_j^\text{out}}, & i\neq j,
		\end{cases}
	\end{align*}
	where $a_{ij}$ is the $ij$-th element of the binary adjacency matrix $A$. 
	It can be easily seen that $w_{ij}^\text{PR}$ satisfies \eqref{eq:defOfWeight} and its weight matrix $W^\text{PR}:=[w_{ij}^\text{PR}]$ is obtained as 
	\begin{align*}
		W^\text{PR}=m I + (1-m) AD^{-1}_\text{out},
	\end{align*}
	where $D_\text{out}$ is a diagonal matrix whose diagonal components are $d_1^\text{out},\cdots,d_N^\text{out}$ in sequence. 
	Since $W^\text{PR}$ is column-stochastic, it has the spectral radius of 1 with the left eigenvector $q=\1_N$. By Lemma \ref{lemPQ}, there exists a positive right eigenvector $p\in\R^N$ for the eigenvalue 1 such that
	\begin{align*}
		W^\text{PR}p=p, \quad \1_N^\top p = 1.
	\end{align*}
	Here, each element $p_i$ of $p$ is called as PageRank score of node $i$ and it represents the relative importance of each node in the network \citep{SB98}. 
	From this, the blended dynamics under PageRank coupling \eqref{eq:prDynamics} is written as, with $s\in\R$,
	\begin{align}\label{eq:prBlendedDynamics1}
		\begin{split}
			s[t+1] = \1_N^\top F\left(t, p s[t]\right) = \sum_{i=1}^{N} f_i\left(t, p_i s[t]\right).
		\end{split}
	\end{align}
		
	By the results in Section~\ref{sect:main}, the $i$-th agent's trajectory over the integer count $t$ is approximated by $p_i s[t]$, which PageRank-scaled solution $s$ of the blended dynamics \eqref{eq:prBlendedDynamics1}. 
	Therefore, if one is interested in solving the PageRank score of each node, then the blended dynamics can be designed to have a stable equilibrium of 1.
	
	In fact, when the network has a large number of agents, these PageRank scores are not easy to be computed in a centralized manner. 
	Thus, the {\em distributed} PageRank algorithms have been proposed in \citep{HI10,HI12,JL14}.
	Unfortunately, most of them commonly assume an initialization. 
	However, when nodes are added to or removed from the network during the process of the algorithm, the whole algorithm must be re-initialized whenever a change occurs in the network, and this is not easy to be achieved in a distributed manner. 
	
	On the contrary, we can design an initialization-free distributed PageRank estimation algorithm by employing the proposed multi-step coupling framework. 
	It can be easily inferred that, if the solution of the blended dynamics simply converges to 1, then every sampled state $x_i[t]$ will approach to its PageRank score $p_i$ under the multi-step coupling of \eqref{eq:prDynamics}. Thus, we first design the blended dynamics which has a stable equilibrium point at 1 as 
	\begin{align}\label{eq:appPR_BlendedDynamics}
		\begin{split}
			s[t+1]
			&=\nu s[t]+(1-\nu),
		\end{split}
	\end{align}
	where $\nu\in(0,1)$ is a design parameter. 
	Since $\nu s[t]+(1-\nu) = \sum_{i=1}^N \left\{ \nu p_i s[t] + (1-\nu)/N \right\}$, we can divide \eqref{eq:appPR_BlendedDynamics} to each node by proposing the following algorithm 
	\begin{align}\label{eq:pageRankDynamics}
		x_i[t_{k+1}]=
		\begin{cases}
			\nu x_i[t_{k}] + \displaystyle \frac{1-\nu}{N}, & \text{if }k=0,\\
			m x_i[t_{k}] + (1-m) \displaystyle \sum_{j\in\NN_i} \frac{x_j[t_{k}]}{d_j^\text{out}}, & \text{otherwise}.
		\end{cases}
	\end{align}
	Indeed, the proposed algorithm has the blended dynamics of \eqref{eq:appPR_BlendedDynamics}.
	As stated in Theorem \ref{thm1} or Corollary \ref{thm2}, the proposed distributed algorithm does not rely on a particular initialization.
	The algorithm \eqref{eq:pageRankDynamics} uses a global information of $N$, but it can be distributively estimated by the result in Section~\ref{subSect:MHCp}.

	\subsection{Distributed Degree Sequence Estimation with Average Coupling}\label{subSect:AvgCp}
	
	Average consensus protocol has been widely utilized in many discrete-time consensus problems including \citep{RS07,WR05}. Thus, in this subsection, we consider a multi-step coupling framework whose coupling dynamics \eqref{eq:couplingDynamics} is the following average consensus protocol
	\begin{align*}
		x_i[t_{k+1}]&=\theta x_i[t_{k}] + \frac{1-\theta}{|\NN_i|} \displaystyle\sum_{j\in\NN_i} x_j[t_{k}],
	\end{align*}
	where $\theta\in(0,1)$ is a parameter which represents weight between its own state and the average of the neighbors. 
	
	Then, the coupling weight $w_{ij}^\text{avg}$ is obtained as
	\begin{align*}
		w_{ij}^\text{avg}=
		\begin{cases}
			\theta, & i=j,\\
			(1-\theta) \displaystyle \frac{a_{ij}}{d_i}, & \text{otherwise},
		\end{cases}
	\end{align*}
	and the weight matrix $W^\text{avg}:=[w_{ij}^\text{avg}]$ is given by
	\begin{align*}
		W^\text{avg}=\theta I + (1-\theta) D^{-1}A,
	\end{align*}
	where $D$ is the diagonal matrix whose diagonal components are $d_1,\ldots,d_N$ sequentially. 
	Note that $w_{ij}^\text{avg}$ satisfies \eqref{eq:defOfWeight} and Assumption~\ref{assW} holds because $W^\text{avg}$ is row-stochastic matrix. 
	Thus, we can choose $p=\1_N$ and find a positive vector $q$ by Lemma \ref{lemPQ} such that
	\begin{align*}
		q^\top W^\text{avg}= q^\top, \quad q^\top \1_N=1.
	\end{align*}
	Now, the blended dynamics is given by
	\begin{align}\label{eq:avgBlendedDynamics1}
		\begin{split}
			s[t+1] = \sum_{i=1}^{N} q_i f_i(t, s[t]). 
		\end{split}
	\end{align}
	Meanwhile, if the network under consideration is undirected, $q$ is easily obtained as $q=(1/d_\text{sum})d$ for $p=\1_N$ where $d_\text{sum}:=\sum_{j=1}^N d_j$ and $d=[d_1,\ldots,d_N]^\top \in \R^N$ because $d^\top D^{-1} A = \1_N^\top A = [d_1^\text{out}, \ldots, d_N^\text{out}]=[d_1,\ldots,d_N]=d^\top$.
	From this, the blended dynamics \eqref{eq:avgBlendedDynamics1} is rewritten as
	\begin{align}\label{eq:avgBlendedDynamics2}
		\begin{split}
			s[t+1]&=\frac{1}{d_\text{sum}} \sum_{i=1}^{N} d_i f_i(t, s[t]).
		\end{split}
	\end{align}
	
	Similarly with Section \ref{subSect:MHCp}, the overall trajectories of $x_i[t]$ are approximately synchronized to the solution of blended dynamics \eqref{eq:avgBlendedDynamics1} (or \eqref{eq:avgBlendedDynamics2}) for sufficiently large $K$. 
	The difference between the doubly-stochastic and row-stochastic weight matrix is that the former has the blended dynamics as the simple average of $f_i$s like \eqref{eq:mhBlendedDynamics}, while the latter has the blended dynamics as the weighted average like \eqref{eq:avgBlendedDynamics1} whose weights $q_i$s are uneven in general.
	
	For undirected graphs, a non-increasing sequence of all degrees is called as {\it degree sequence} \citep{RD17}. 
	Since the degree sequence does not uniquely identify a graph, there has been much attention to obtain information of the graph structure from the given degree sequence. 
	For example, \citet{FV05} realized the given degree sequence by a simple graph (realization problem) and \citet{FH14} estimated the number of graphs with the given degree sequence (graph enumeration). 
	If each agent can predict possible structures of the network with the degree sequence, it could obtain global information such as the algebraic connectivity. 
	To achieve this, a distributed algorithm to estimate the degree sequence is required, and we proposed one implementation by employing the proposed framework as follows.
	
	In the proposed algorithm, we additionally assume that each agent knows the network size $N$ and has its unique id. Again, $N$ can be distributively estimated using the application example in Section \ref{subSect:MHCp}. 
	Moreover, the assumption on the unique id for each agent is quite natural in the sense that, in practice, every communication device has its own identifier such as mac address.
	Let $\XX_i\in\Z$ be the id of the agent $i$ and suppose $1<\XX_i$ for all $i\in\NN$ without loss of generality.
	
	Under the above assumptions, an arbitrary $i$-th agent runs the following dynamics
	\begin{align*}
		x_i[t_{k+1}] = 
		\begin{cases}
			\left( 1 - 1/d_i \right) x_i[t_{k}]+ {N}^{\XX_i}, & \text{if }k=0,\\
			\sum_{j\in\NN_i\cup\{i\}} w_{ij}^\text{avg} x_j[t_{k}], & \text{otherwise},
		\end{cases}
	\end{align*}
	where $x_i\in\R$ is the state. 
	From \eqref{eq:avgBlendedDynamics2}, the blended dynamics is obtained as 
	\begin{align*}
		s[t+1]&=\frac{1}{d_\text{sum}} \left\{ \sum_{i=1}^N d_i \left( 1- \frac{1}{d_i} \right) s[t] + \sum_{i=1}^N d_i {N}^{\XX_i} \right\}\\
		&=\left( 1-\frac{N}{d_\text{sum}} \right) s[t] + \frac{1}{d_\text{sum}} \sum_{i=1}^N d_i {N}^{\XX_i}.
	\end{align*}
	For any connected network, $d_\text{sum} \ge N$, and this guarantees the contraction stability of the above blended dynamics. 
	In addition, its equilibrium point $s^*$ is obtained as
	\begin{align*}
		s^* = \frac{1}{N} \sum_{i=1}^N d_i {N}^{\XX_i} = \sum_{i=1}^N d_i {N}^{\XX_i-1}.
	\end{align*}
	Meanwhile, it is easily seen that $d_i < N$ for all $i \in \NN$ because the network has no self-connection. 
	Thus, the number $s^*$ can be regarded as a representation of the numeral system with the base ${N}$:
	\begin{align}\label{eq:numeralSystem}
		[s^*]_N = [\delta_{(B-1)} \delta_{(B-2)} \cdots \delta_1 \delta_0]_{N},
	\end{align}
	where $B=\max_{i\in\NN} \XX_i$ and $\delta_b$ represents the $(b+1)$-th right-most digit such that $\delta_b = d_i$ if $b=\XX_i-1$ and $\delta_b=0$ otherwise. 
	
	As a result, since every state $x_i[t]$ is approximately synchronized to $s^*$ of \eqref{eq:numeralSystem}, each agent can estimate the degree sequence by removing zero digits in the $N$-base numeral representation of its state and ordering the rest digits. 
	Here, we suppose that $K$ is properly chosen such that the synchronization error $\epsilon<1$ because this guarantees that the error does not affect even in the right-most digit $\delta_0$.

	\section{Conclusion} \label{sect:conclusion}
	
	In this paper, we introduced a discrete-time blended dynamics theorem, which inherits all the benefits of the continuous-time blended dynamics theorems in \citep{JL20}.
	This was achieved by the proposed multi-step coupling in the multi-agent algorithm \eqref{eq:maineq}. 
	The proposed approach does not require stability of individual agents as long as the blended dynamics is stable, the plug-and-play operation is easily achieved.
	To illustrate utility of the proposed method as a design tool, three application examples are included.

	\bibliographystyle{pre-automatica}
	\bibliography{root}
	
	\appendix

	\section{Proof of Lemma~\ref{lemNormIneq} } \label{apdx:lemNormIneq}
	
	Before proving Lemma~\ref{lemNormIneq}, we first claim that, for each $t\in\Z$ and $s_1,s_2\in\R^n$, there exists $\tilde{s}\in\R^n$ in the line connecting $s_1$ and $s_2$ such that
	\begin{gather}\label{eq:contQuadIneq}
		\begin{split}
			\{f_s(t,s_2)-f_s(t,s_1)\}^\top H^2 \{f_s(t,s_2)-f_s(t,s_1)\} \\
			\le (s_2-s_1)^\top \frac{\partial f_s}{\partial s}(t,\tilde{s})^\top H^2 \frac{\partial f_s}{\partial s} (t,\tilde{s}) (s_2-s_1).
		\end{split}
	\end{gather}
	It can be proved by the mean-value theorem with a trick. Let the left-hand side of the equality \eqref{eq:contQuadIneq} as $\Delta(t,s_1,s_2)$ for convenience. With a variable $c\in\R$, define
	\begin{align*}
		&g_{t,s_1,s_2}(c) := \{f_s(t,s_2) - f_s(t,s_1)\}^\top H^2 \\
		&\qquad\qquad\quad \times \{f_s(t,cs_2 + (1-c)s_1)-f_s(t,s_1)\}.
	\end{align*}
	Since the function $g_{t,s_1,s_2}:\R \rightarrow \R$ is continuously differentiable, by the mean-value theorem, there exists $\tilde{c}\in[0,1]$ such that $g_{t,s_1,s_2}(1)-g_{t,s_1,s_2}(0)=g_{t,s_1,s_2}'(\tilde{c})(1-0)$, which is equivalent to
	\begin{align}\label{eq:meanValueTheorem}
		\begin{split}
			\Delta(t,s_1,s_2) &= \{f_s(t,s_2) - f_s(t,s_1)\}^\top H^2 \\
			&\qquad\qquad\qquad \times\frac{\partial f_s}{\partial s}(t,\tilde{s})(s_2-s_1)
		\end{split}
	\end{align}
	where $\tilde{s}=\tilde{c}s_2 + (1-\tilde{c})s_1$.
	Using this, we have
	\begin{align*}
		&\Delta(t,s_1,s_2) = \| H \{f_s(t,s_2)-f_s(t,s_1)\} \|^2 \\
		&\le \| H \{ f_s(t,s_2) - f_s(t,s_1) \} \| \bigg\| H \frac{\partial f_s}{\partial s}(t,\tilde{s})(s_2-s_1) \bigg\|,
	\end{align*}
	which in turn implies $\|H \{f_s(t,s_2) - f_s(t,s_1)\}\| \le \|H (\partial f_s/\partial s) (t,\tilde{s})(s_2-s_1)\|$. 
	From this, the claim \eqref{eq:contQuadIneq} is justified.
	Finally, it follows from Assumption~\ref{assStability} that
	\begin{align*}
		&\| H \{f_s(t,s_2)-f_s(t,s_1)\} \|^2 = \Delta(t,s_1,s_2) \\
		&\le (s_2-s_1)^\top \frac{\partial f_s}{\partial s}(t,\tilde{s})^\top H^2 \frac{\partial f_s}{\partial s} (t,\tilde{s}) (s_2-s_1)\\
		&\le \gamma (s_2 - s_1)^\top H^2 (s_2 - s_1) = \gamma \| H (s_2 - s_1)\|^2,
	\end{align*}
	which proves Lemma~\ref{lemNormIneq}.

	\section{Proof of Theorem~\ref{thm1}} \label{apdx:thm}	
	
	The proof begins with the claim that the solution $s[t]$ of the blended dynamics is bounded by Assumptions \ref{assStability}--\ref{assFi}.	
	
	\begin{lem1}\label{lemBoundOfBD}
		Under Assumption~\ref{assStability}, the solutions of the blended dynamics \eqref{eq:blendedDynamics0}, initiated at time $t=t^0$, are bounded as
		\begin{align}\label{eq:boundOfW0}
			\|Hs[t]\| \le \sqrt{\gamma}^{t-t^0} \|Hs[t^0]\| + \frac{\sup_{\tau \ge t^0} \|H f_s(\tau,\0_n)\|}{1-\sqrt{\gamma}}.
		\end{align}
		Additionally, with Assumption \ref{assFi}, we have 
		\begin{align}\label{eq:lemBound0}
			\begin{split}
				\limsup_{t\rightarrow\infty} \|s[t]\| &\le \frac{\sqrt{N} \|q\| \|H^{-1}\|\|H\| M(0)}{1-\sqrt{\gamma}} =: M_s
			\end{split}
		\end{align}
	\end{lem1}
	
	\begin{pf}
		We prove \eqref{eq:boundOfW0} by showing that
		\begin{align}\label{eq:compInequality}
			\|H s[t]\|\le w[t], \quad \forall t \ge t^0
		\end{align}
		where $w \in \R$ is the solution of
		\begin{align}\label{eq:compDynamics}
			w[t+1] = \sqrt{\gamma} w[t] + \|H f_s(t,\0_n)\|, \quad w[t^0] =\|Hs[t^0]\|.
		\end{align}
		Indeed, since $\|H s[t^0]\| \le w[t^0]$, let us suppose $\|H s[\tau]\|\le w[\tau]$ for some integer $\tau \ge t^0$.
		Then, 
		\begin{align*}
			&\|H s[\tau+1]\| = \|H \{f_s(\tau,s[\tau]) - f_s(\tau,\0_n) + f_s(\tau,\0_n)\}\| \\
			&\qquad \le \|H \{f_s(\tau,s[\tau])-f_s(\tau,\0_n)\}\| + \|H f_s(\tau,\0_n)\| \\
			&\qquad \le \sqrt{\gamma} \|H s[\tau]\| + \|H f_s(\tau,\0_n)\| \\
			&\qquad \le \sqrt{\gamma} w[\tau] + \|H f_s(\tau,\0_n)\| = w[\tau+1],
		\end{align*}
		where the second inequality comes from Lemma~\ref{lemNormIneq}.
		This justifies \eqref{eq:compInequality}.
		Meanwhile, the solution $w$ of \eqref{eq:compDynamics} is given by
		\begin{align*}
			w[t] = \sqrt{\gamma}^{t-t^0} w[t^0] + \sum_{\tau=t^0}^{t-1} \sqrt{\gamma}^{t-\tau-1} \|H f_s(\tau,\0_n)\|,
		\end{align*}
		which yields \eqref{eq:boundOfW0}.
		
		Now, with Assumption \ref{assFi}, it follows from \eqref{eq:boundOfW0} that
		\begin{align*}
			&\limsup_{t \to \infty} \|s[t]\| \le \|H^{-1}\| \limsup_{t \to \infty} \|H s[t]\| \\
			&\quad \le \|H^{-1}\| \|H\| \frac{\sup_{\tau \ge t^0} \|q_{\otimes n}^\top F(t,p_{\otimes n} \0_n)\|}{1-\sqrt{\gamma}} \\
			&\quad \le \|H^{-1}\| \|H\| \frac{\|q\| \sqrt{N} M(0)}{1 - \sqrt{\gamma}}
		\end{align*}
		which completes the proof. 
		$\hfill\blacksquare$
	\end{pf}
		
	Now, we analyze the behavior of \eqref{eq:xiDynamics} with \eqref{eq:blendedDynamics}, which describes the evolution of the overall system at every integer time $t$.
	For this, we introduce a Lyapunov function 
	$$V = \| H (\xi_1 - s) \| + \eta \|\tilde{\xi}\|$$
	where $\eta > L \|q\| \|R\| \|H\|/\sqrt{\gamma}$.
	Then,
	\begin{align*}
		&V[t+1] = \| H (\xi_1[t+1] - s[t+1]) \| + \eta\|\tilde{\xi}[t+1]\| \\
		&= \big\| H \{ q_{\otimes n}^\top F(t, p_{\otimes n}\xi_1[t]) - q_{\otimes n}^\top F(t, p_{\otimes n}s[t]) \\ 
		&\quad \;\; + q_{\otimes n}^\top F(t, p_{\otimes n}\xi_1[t] + R_{\otimes n}\tilde{\xi}[t]) - q_{\otimes n}^\top F(t, p_{\otimes n}\xi_1[t]) \} \big\| \\
		&\quad + \eta \big\| (\Lambda^{K-1} Z^\top)_{\otimes n} \{ F(t, p_{\otimes n}s[t]) \\
		&\quad \;\; + F(t, p_{\otimes n}\xi_1[t]) - F(t, p_{\otimes n}s[t]) \\
		&\quad \;\; + F(t, p_{\otimes n}\xi_1[t] + R_{\otimes n}\tilde{\xi}[t]) - F(t, p_{\otimes n}\xi_1[t]) \} \big\| \\
		&\le \big\| H \{q_{\otimes n}^\top F(t, p_{\otimes n}\xi_1[t]) - q_{\otimes n}^\top F(t, p_{\otimes n}s[t])\} \big\| \\
		&\quad + \big\| H q_{\otimes n}^\top \{ F(t, p_{\otimes n}\xi_1[t] + R_{\otimes n}\tilde{\xi}[t]) - F(t, p_{\otimes n}\xi_1[t]) \} \big\| \\
		&\quad + |\lambda_2|^{K-1}\eta \Big\{ \| Z_{\otimes n}^\top \{ F(t, p_{\otimes n}\xi_1[t]) - F(t, p_{\otimes n}s[t]) \} \| \\
		&\quad \;\; + \| Z_{\otimes n}^\top \{ F(t, p_{\otimes n}\xi_1[t] + 	R_{\otimes n}\tilde{\xi}[t]) - F(t, p_{\otimes n}\xi_1[t]) \} \| \\
		&\quad \;\; + \| Z_{\otimes n}^\top F(t, p_{\otimes n}s[t])\| \Big\}.
	\end{align*}
	The above inequality can be simplified by the following properties:
	\begin{itemize}
		\item by Lemma~\ref{lemNormIneq},
		\begin{align*}
			&\| H \{ q_{\otimes n}^\top F(t, p_{\otimes n}\xi_1[t]) - q_{\otimes n}^\top F(t, p_{\otimes n}s[t]) \} \|\\
			&\quad = \| H \{f_s(t, \xi_1[t]) - f_s(t, s[t])\} \| \\
			&\quad \le \sqrt{\gamma} \| H (\xi_1[t] - s[t]) \|
		\end{align*}
		\item by Assumption \ref{assFi},
		\begin{align*}
			&\| F(t,p_{\otimes n} \xi_1[t] + R_{\otimes n} \tilde{\xi}[t]) - F(t,p_{\otimes n} \xi_1[t]) \| \\
			&\le \left(\sum_{i=1}^N \| f_i(p_i \xi_1[t] + (R_i\otimes I_n) \tilde{\xi}[t]) - f_i(p_i \xi_1[t]) \|^2 \right)^{1/2} \\
			&\le \left(\sum_{i=1}^N L^2 \| (R_i \otimes I_n) \tilde{\xi}[t] \|^2 \right)^{1/2}\\
			&= L \| R_{\otimes n} \tilde{\xi}[t] \| \le L \|R\| \|\tilde{\xi}[t]\|,
		\end{align*}
		where $R_i$ is the $i$-th row of $R$, and similarly
		\begin{align*}
			&\| F(t, p_{\otimes n} \xi_1[t]) - F(t,p_{\otimes n} s[t]) \|\\
			&\le L \|p\| \|\xi_1[t] - s[t]\|= L \|p\| \| H^{-1} H (\xi_1[t] - s[t])\| \\
			&\le L \|p\| \|H^{-1}\| \|H (\xi_1[t] - s[t])\|.
		\end{align*}
	\end{itemize}
	Using the above three inequalities, we have that 
	\begin{align*}
		V[t+1] &\le \sqrt{\gamma} \| H (\xi_1[t] - s[t]) \| + \|H\| \|q\| L \|R\| \|\tilde{\xi}[t]\|\\
		&\quad + |\lambda_2|^{K-1} \eta \|Z\| L \|p\| \|H^{-1}\| \| H (\xi_1[t] - s[t]) \|\\
		&\quad + |\lambda_2|^{K-1} \eta \|Z\| L \|R\| \|\tilde{\xi}[t]\| \\
		&\quad + |\lambda_2|^{K-1} \eta \|Z\| \| F(t, p_{\otimes n}s[t]) \| \\
		&\le \sqrt{\gamma} V[t] + |\lambda_2|^{K-1} \eta {L} \|Z\| M_1 V[t] \\
		&\quad + |\lambda_2|^{K-1} \eta \|Z\| \|F(t, p_{\otimes n}s[t])\|,
	\end{align*}
	where $M_1 := \max \{ \|p\| \|H^{-1}\|, \|R\|/\eta \}$.
	
	For the given $\epsilon$, let $K^\mathrm{min}$ be a positive integer such that 
	\begin{gather}
		|\lambda_2|^{{K}^\mathrm{min}} \eta {L} M_1 \|Z\| \le \frac{1-\sqrt{\gamma}}{2} \label{eq:kmin1} \\
		|\lambda_2|^{{K}^\mathrm{min}} \frac{2 \eta M_1 M(\|p\| M_s) \sqrt{N} \|Z\|}{1-\sqrt{\gamma}} \le \frac{\epsilon}{2} . \label{eq:kmin2}
	\end{gather}
	
	Then, for all $K > {K}^\mathrm{min}$,
	\begin{multline}\label{eq:ultBound}
		V[t+1] - V[t] \le -\frac{(1-\sqrt{\gamma})}{2} V[t] \\
		+ |\lambda_2|^{K-1} \eta \|Z\| \|F(t, p_{\otimes n}s[t])\|.
	\end{multline}
	By Assumption \ref{assFi} and by Lemma~\ref{lemBoundOfBD},
	\begin{align*}
		\limsup_{t \to \infty} \|F(t, p_{\otimes n}s[t])\| &\le \sqrt{N} M(\|p\| \limsup_{t \to \infty} \| s[t] \|) \\
		&\le \sqrt{N} M\left(\|p\| M_s\right)
	\end{align*}
	Using this and \eqref{eq:ultBound}, the ultimate bound of $V$ is obtained as
	\begin{align}\label{eq:LyapunovBound}
		\limsup_{t \to \infty} V[t] \le |\lambda_2|^{K-1}\frac{2 \eta \|Z\|}{1-\sqrt{\gamma}} \sqrt{N} M(\|p\| M_s).
	\end{align}
	Therefore, for each agent $i\in\mathcal{N}$ and $K > {K}^\mathrm{min}$,
	\begin{align}\label{eq:errorBound}
		\begin{split}
			&\limsup_{t\rightarrow\infty} \| {x}_i[t]-p_i s[t] \| \\
			&= \limsup_{t\rightarrow\infty} \left\| p_i (\xi_1[t] - s[t]) + (R_i\otimes I_n) \tilde{\xi}[t] \right\| \\
			&\le \max\left\{\| p\| \|H^{-1}\|,\frac{\| R\|}{\eta}\right\} \limsup_{t\rightarrow\infty} V[t]\\
			&\le |\lambda_2|^{K-1}\frac{2 \eta \|Z\|}{1-\sqrt{\gamma}} \sqrt{N} M(\|p\| M_s) M_1
			\le \epsilon
		\end{split}
	\end{align} 
	where we used $\bar x = p_{\otimes n}\xi_1+R_{\otimes n}\tilde{\xi}$.
	This completes the proof for \eqref{eq:thm1Eq1} of Theorem~\ref{thm1}.	
	
	Now, in order to inspect the behavior of the system over the fractional time, let us apply the transformation of \eqref{eq:cordTrans} to \eqref{eq:realoverall}, which yields, for $k = 1, \cdots, K-1$,
	\begin{align}\label{eq:xixi}
		\begin{split}
			\xi_1[t_k] &= q_{\otimes n}^\top W_{\otimes n}^{k-1} F(t_0, \bar x[t_0]) \\
			&= q_{\otimes n}^\top F(t_0, \bar x[t_0]) = q_{\otimes n}^\top \bar x[(t+1)_0] \\
			&= \xi_1[(t+1)_0]
		\end{split}
	\end{align}
	in which, the third equality can also be seen from \eqref{eq:periodicDynamics}.
	Similarly, for $k = 1, \cdots, K-1$,
	\begin{align*}
		\tilde{\xi}[t_k] &= Z_{\otimes n}^\top W_{\otimes n}^{k-1} F(t_0, \bar x[t_0]) \\
		&= (\Lambda^{k-1} Z^\top)_{\otimes n} F(t_0,\bar x[t_0]) \\
		&= \Lambda^{k-K}_{\otimes n} (\Lambda^{K-1}Z^\top)_{\otimes n} F(t_0,\bar x[t_0]) \\
		&= \Lambda^{k-K}_{\otimes n} Z^\top_{\otimes n} W^{K-1}_{\otimes n} F(t_0,\bar x[t_0]) \\
		&= (\Lambda^{-1})_{\otimes n}^{K-k} \tilde \xi[(t+1)_0].
	\end{align*}
	Hence,
	\begin{equation}\label{eq:tildexitildexi}
		\|\tilde{\xi}[t_k]\| \le \frac{\|\tilde \xi[(t+1)_0]\|}{|\lambda_N|^{K-k}}
	\end{equation}
	for each $k=1,\cdots,K-1$.
	
	On the other hand, by \eqref{eq:LyapunovBound} and \eqref{eq:kmin2}, we have
	\begin{gather*}
		\limsup_{t\to\infty} \max\{ \|H (\xi_1[t]-s[t])\|, \eta\|\tilde\xi[t]\| \} \\
		\qquad \le \limsup_{t\to\infty} V[t] \le \frac{\epsilon}{2 M_1}.
	\end{gather*}
	Therefore, for each $k=1,\cdots,K-1$,
	\begin{align*}
		&\limsup_{t\to\infty} \|x_i[t_k] - p_i s[t+1]\| \\
		&= \limsup_{t \to \infty} \| p_i(\xi_1[t_k] - s[t+1]) + (R_i \otimes I_n) \tilde \xi[t_k]\| \\
		&\le \limsup_{t \to \infty} \|p\| \|H^{-1}\| \| H(\xi_1[t_k]-s[t+1]) \| + \frac{\|R\|}{\eta} \eta \|\tilde \xi[t_k]\| \\
		&\le \limsup_{t \to \infty} \|p\| \|H^{-1}\| \| H(\xi_1[t+1]-s[t+1]) \| \\
		&\qquad \qquad \qquad \qquad \qquad \qquad \qquad + \frac{\|R\|}{\eta}\frac{\eta \|\tilde \xi[t+1]\|}{|\lambda_N|^{K-k}} \\
		&\le \frac{\|p\| \|H^{-1}\|}{M_1} \frac{\epsilon}{2} + \frac{\|R\|/\eta}{M_1} \frac{\epsilon}{2 |\lambda_N|^{K-k}} \\
		&\le \frac{\epsilon}{2} \left( 1 + \frac{1}{|\lambda_N|^{K-k}} \right) ,
	\end{align*}
	which completes the proof.

	\section{Proof of Corollary~\ref{thm2} } \label{apdx:thm2}
	
	In this proof, we show that, after $K-1$ times execution of \eqref{eq:couplingDynamics}, the solution of the overall system from the initial condition enters a positively invariant set, in which \eqref{eq:corineq1} holds.
	For this, let us first construct a few sets as
	\begin{align*}
		C_1^s &= \{q_{\otimes n}^\top F(0,\bar x): x_i \in C, i \in \NN \} \subset \mathbb{R}^n, \\
		C_{t+1}^s &= \{q_{\otimes n}^\top F(t,p_{\otimes n}s) : s \in C_{t}^s\} \subset \mathbb{R}^n, \; \forall t \ge 1, \\
		C_\infty^s &= \bigcup_{t=1}^\infty C_t^s \subset \mathbb{R}^n
	\end{align*}
	in which, $C_t^s$ is the set of all possible $s[t]$ for each $t \ge 1$, which is bounded.
	The set $C_\infty^s$ is also bounded by Lemma~\ref{lemBoundOfBD}.
	Now, for the overall state $\bar x$, consider two more sets:
	\begin{align*}
		C' &= \{ \bar x : \| H (\xi_1 - s) \| \le \epsilon_0, s \in C_\infty^s, \|\tilde{\xi}\| \le \delta \} \cup C^N \\
		&\subset \mathbb{R}^{nN}, \\
		\bar{C} &= \{(\bar x,s) \in C' \times C_\infty^s : \| H (\xi_1 - s) \| \le \epsilon_0, \|\tilde{\xi}\| \le \delta \} \\
		&\subset \mathbb{R}^{nN} \times \R^{n},
	\end{align*}
	where $C^N$ is $N$-ary Cartesian power of $C$, i.e., $C^N = \underbrace{C \times C \times \cdots \times C}_{N\text{-times}}$ and
	\begin{align*}
		\epsilon_0 &:= \frac{\epsilon}{2\max\{\|p\| \|H^{-1}\|, \|R\|\}}, \\
		\delta &:= \min\left\{1, \frac{1-\sqrt{\gamma}}{L \|q\| \|R\| \|H\|}\right\} \epsilon_0.
	\end{align*}
	Now, pick $K^\mathrm{min}$ such that
	$$|\lambda_2|^{K^\mathrm{min}} \|Z\| \sup_{\bar x \in C'} \|F(t, \bar x)\| \le \delta, \quad \forall t \ge 0.$$
	We claim that the set $\bar C$ is positively invariant for any $K > K^\mathrm{min}$.
	To see this, suppose that, for any $\tau \ge 1$, $s[\tau] \in C_\tau^s \subset C_\infty^s$, $\|\tilde{\xi}[\tau]\| \le \delta$, and $\| H (\xi_1[\tau] - s[\tau]) \| \le \epsilon_0$, so that $(\bar x[\tau],s[\tau]) \in \bar C$.
	Then, it follows that $s[\tau+1] \in C_{\tau+1}^s \subset C_\infty^s$, $\|\tilde{\xi}[\tau+1]\| \le |\lambda_2|^{K-1} \|Z\| \|F(\tau, \bar x[\tau])\| \| \le \delta$, and
	\begin{align*}
		&\| H (\xi_1[\tau+1] - s[\tau+1]) \| \\
		&\le \| H \{ \xi_1[\tau+1] - q_{\otimes n}^\top F(\tau, p_{\otimes n}\xi_1[\tau]) \} \| \\
		&\quad + \| H \{ q_{\otimes n}^\top F(\tau, p_{\otimes n}\xi_1[\tau]) - s[\tau+1]) \} \| \\
		&\le L \|q\| \|R\| \|H\| \delta + \| H \{f_s(\tau, \xi_1[\tau]) - f_s(\tau, s[\tau])\} \| \\
		&\le L \|q\| \|R\| \|H\| \delta + \sqrt{\gamma} \|H (\xi_1[\tau] - s[\tau]) \| \\
		&\le \epsilon_0
	\end{align*}
	in which, we used
	\begin{align*}
		&\| \xi_1[\tau+1] - q_{\otimes n}^\top F(\tau, p_{\otimes n}\xi_1[\tau]) \|\\
		&= \| q_{\otimes n}^\top \{ F(\tau, p_{\otimes n}\xi_1[\tau] + R_{\otimes n}\tilde{\xi}[\tau]) - F(\tau, p_{\otimes n}\xi_1[\tau]) \} \| \\
		&\le L\|q\|\|R\| \|\tilde{\xi}[\tau]\| \le L\|q\|\|R\|\delta.
	\end{align*}
	On the other hand, the set $\bar C$ is reached within $K-1$ executions of \eqref{eq:couplingDynamics} from the initial time $0_0$.
	Indeed, $s[1] = q_{\otimes n}^\top F(0, \bar x[0]) \in C_1^s \subset C_\infty^s$, $\|\tilde{\xi}[1]\| = \| (\Lambda^{K-1} Z^\top)_{\otimes n} F(0, \bar x[0]) \| \le |\lambda_2|^{K-1} \|Z\| \|F(0,\bar x[0])\| \le \delta$, and $\| H (\xi_1[1] - s[1]) \| = \| H \{ q_{\otimes n}^\top W_{\otimes n}^{K-1} F(0, \bar x[0]) - q_{\otimes n}^\top F(0, \bar x[0]) \} \| = 0$.
	Therefore, $(\bar x[t],s[t])$ remains in $\bar C$ for all $t \ge 1$.
	
	Finally, it is seen that, in the set $\bar C$,
	\begin{align*}
		&\| x_i[t]-p_i s[t] \| = \| p_i (\xi_1[t] - s[t]) + (R_i\otimes I_n) \tilde{\xi}[t] \| \\
		&\le \max\left\{\| p\| \|H^{-1}\|, \|R\| \right\} ( \| H (\xi_1[t] - s[t]) \| + \|\tilde{\xi}[t]\|)\\
		&\le \epsilon,
	\end{align*}
	which completes the proof of \eqref{eq:corineq1}.	
	
	To show \eqref{eq:corineq2}, we note that \eqref{eq:xixi} and \eqref{eq:tildexitildexi} still hold.
	Therefore, for all $k = 1,\cdots, K-1$, $i\in\NN$, and $t \ge 1$, 
	\begin{align*}
		&\| x_i[t_k] - p_i s[t+1] \| \\
		&= \| p_i (\xi_1[t_k] - s[t+1]) + (R_i\otimes I_n) \tilde{\xi}[t_k] \| \\
		&\le \max\left\{\| p\| \|H^{-1}\|, \| R\| \right\} \\
		&\quad \times ( \| H (\xi_1[t_k] - s[t+1]) \| + \|\tilde{\xi}[t_k]\|) \\
		&\le \max\left\{\| p\| \|H^{-1}\|, \| R\| \right\} \\
		&\quad \times ( \| H (\xi_1[t+1] - s[t+1]) \| + \|\tilde{\xi}[t+1]\|/|\lambda_N|^{K-k}) \\
		&\le \max\left\{\| p\| \|H^{-1}\|, \| R\| \right\} \left( \epsilon_0 + \frac{\delta}{|\lambda_N|^{K-k}} \right) \\
		&\le \frac{\epsilon}{2} \left( 1 + \frac{1}{|\lambda_N|^{K-k}} \right),
	\end{align*}
	which completes the proof of \eqref{eq:corineq2}.	
	
\end{document}